\documentclass{amsart}

\newcommand{\pf}{\noindent {\bf Proof. }}

\renewcommand{\r}{\rightarrow}
\newcommand{\e}{\epsilon}
\newcommand{\ep}{$\Box$}
\renewcommand{\dfrac}{\displaystyle\frac}

\newtheorem{lemma}{Lemma}[section]
\newtheorem{proposition}[lemma]{Proposition}
\newtheorem{theorem}[lemma]{Theorem}

\begin{document}

\title{Existence of traveling waves for the nonlocal Burgers equation}

\author{Adam J.J. Chmaj}
\address{Department of Integral Equations \\
Faculty of Mathematics and Information Science \\
Warsaw University of Technology\\ Pl. Politechniki 1 \\
00-661 Warsaw \\ Poland} \email{A.Chmaj@mini.pw.edu.pl}
\date{27.03.2006}

\begin{abstract}
We study the equation $u_t +uu_x +u-K*u=0$ in the case of an
arbitrary $K\geq 0$, which is a generalization of a model for
radiating gas, in which $K(y)=\frac{1}{2}e^{-|y|}$. Using a monotone
iteration scheme argument we establish the existence of traveling
waves, which gives a solution to an open question raised by Denis
Serre \cite{s2}.
\end{abstract}

\maketitle

\section{Introduction}
\setcounter{equation}{0}

The system of equations
\begin{equation}
\left\{ \begin{array}{ll} u_t +uu_x +q_x =0 ,\\
-q_{xx} +q+u_x =0, \end{array} \right. \label{rg}
\end{equation}
for $x\in R$ and $t\geq 0$ was derived by K. Hamer in $1971$
\cite[eqs. (4.16), (4.19)]{h} (see also \cite{r}) as a third-order
approximation of the full system describing the motion of radiating
gas in thermo-nonequilibrium (the second-order approximation there
gives the viscous Burgers equation $u_t +uu_x =u_{xx}$, and the
first-order approximation gives the inviscid Burgers equation $u_t
+uu_x =0$). The system (\ref{rg}) is formally equivalent to
\begin{equation}
u_t +uu_x +u- \frac{1}{2} \int_R e^{-|x-y|}u(y,t)dy =0, \label{gf}
\end{equation}
and can be considered a special case of the integrodifferential
equation
\begin{equation}
u_t +uu_x +u-K*u =0, \label{nb}
\end{equation}
where $K*u(x,t)\equiv \int_R K(x-y)u(y,t)dy$, $K$ is even and
$\int_R K=1$. Nonlinear equations with a nonlocal operator being
considered instead of the Laplacian have a long history and have
recently been gaining increasing popularity, due to superior
physical qualities of such models and at the same time often
ingenious mathematical solutions. In fact, when in $1893$ J.D. van
der Waals derived his variational thermodynamic theory of binary
phase transitions, he introduced a mean field $K$, formally expanded
$K*w-w$ into power series: $\int_R K(y)(w(z+y)-w(z))dy=c_1
w''(z)+c_2 w^{iv}+ \ldots$, where $c_1 =\frac{1}{2} \int_R y^2
K(y)dy$, $c_2 =\frac{1}{24} \int_R y^4 K(y)dy$, etc., and used just
the first term in this expansion \cite{w}. He also investigated his
nonlocal model with the Green's function $K(y)=\frac{1}{2}e^{-|y|}$
\cite{w}.

The Hamer equation (\ref{gf}) was studied in \cite{h,st,kn1,kn2,kt}
and the more general nonlocal Burgers equation (\ref{nb}) in
\cite{lt,s1,s2}. In particular, it was discovered that the solution
of the Cauchy problem for (\ref{nb}) with smooth initial data is
either smooth for all times or becomes discontinuous in finite time.
Here we focus on traveling wave solutions $U(x-st)$ of (\ref{nb}).
To account for possible shock discontinuities, a traveling wave is
defined as a weak solution, i.e., a function $U$ which satisfies
\begin{equation}
\int_R \bigl[ (\frac{1}{2}U^2 -sU)\phi' +(K*U-U)\phi \bigr] =0
\label{wf}
\end{equation}
for an arbitrary $\phi \in C^{\infty}_0$. In \cite{kn1} the authors
obtained the following result (see also a partial version in the
earlier \cite{st}).
\begin{proposition}
\cite{kn1} For every pair $u_- >u_+$ there exists a solution
$U(x-st)$ of (\ref{wf}) with $K(y)=\frac{1}{2}e^{-|y|}$,
$U(-\infty)=u_-$, $U(+\infty)=u_+$ and $s=\frac{1}{2} (u_- +u_+ )$.
Moreover, (a) When $u_- -u_+
>\sqrt{2}$, $U$ is continuous except for one point, (b) When
$u_- -u_+ \leq \sqrt{2}$, $U$ is $C^1$. (c) When $u_- -u_+ \leq
\frac{2\sqrt{2k}}{k+1}$, $U$ is $C^k$.
\end{proposition}
Subsequently, Denis Serre raised the question whether a similar
result can be established for (\ref{nb}) \cite[Open Problem 1, p.
33]{s2}. In Section 2 we answer it in the affirmative, though we
fall short of providing an exact regularity criterion of the
profiles as in Proposition 1.1.

\section{Existence of traveling waves}
\setcounter{equation}{0}

\begin{theorem}
Let $K\geq 0$, $K\in W^{1,1}(R)$, $K$ nonincreasing on $(0,\infty)$,
$\int_R y^2 K(y)dy$ \\$< \infty$. For every pair $u_- >u_+$, there
exists a solution $U(x-st)$ of (\ref{wf}) such that $U' (x) <0$ for
$x \neq 0$, $U(-\infty)=u_-$, $U(+\infty)=u_+$ and $s=\frac{1}{2}
(u_- +u_+ )$. Moreover, if $u_- -u_+ >4 \int_R |y|K(y)dy$, then $U$
is discontinuous at one point.
\end{theorem}

\pf Proposition 1.1 was proved by a phase-plane analysis \cite{kn1}.
Here we argue in an indirect way as we construct a monotone
iteration scheme and pass to the limit. This is a standard elliptic
technique, sometimes called the sub-supersolution method, whose use
in nonlocal problems dates to \cite{ar} (see also the many follow-up
papers on traveling waves in various nonlocal monostable models). In
an early and unpublished version of \cite{bfrw} the authors used it
to construct continuous and discontinuous stationary waves of the
nonlocal bistable equation \cite{bfr}. A similar argument appeared
in \cite{cr} for the bounded interval case.

The main difficulty is to set up the difference scheme in a way that
the sequence of monotone solutions is monotone. The whole sequence
then has a limit so that one can pass to the limit on both sides of
the scheme. This procedure would not work if only a subsequence was
convergent. To ensure the limit is not trivial, one also needs to
find an appropriate subsolution.

It suffices to show that for any constant $u_c >0$,
\begin{equation}
uu'=K*u -u \label{e}
\end{equation}
admits a monotone and odd solution $u$ such that $u(-\infty)=u_c$
and $u(+\infty)=-u_c$. If we let $U=u+s$, where $u_c
=\frac{1}{2}(u_- -u_+ )$ and $s=\frac{1}{2} (u_- +u_+ )$ (the
Rankine-Hugoniot condition), then, since $\int_R K=1$, $U$ satisfies
\[ U' (U-s)=K*U-U, \]
$U(-\infty)=u_-$ and $U(+\infty)=u_+$.

We will see that it suffices to study (\ref{e}) on $(-\infty,0)$. We
find it convenient first to find a supersolution and a subsolution
for (\ref{e}). Let
\[ u_0 (x)\equiv \left\{ \begin{array}{ll} u_c , & x<0 ,\\ 0,&x=0 ,\\ -u_c
,&x>0 .\end{array} \right. \] Then it is easily seen that
\[ u_0 + u_0 u_0' \geq K*u_0 \]
on $(-\infty,0)$, thus $u_0$ is a supersolution there.

Let \[ s(x)\equiv -\frac{2u_c}{\pi} {\rm tan}^{-1} (\e x).\] We
investigate the function
\[ g(x,\e)\equiv \frac{-\int_R K(x-y)({\rm tan}^{-1} (\e y)-{\rm tan}^{-1}
(\e x))dy}{\dfrac{2u_c}{\pi}{\rm tan}^{-1} (\e x) \dfrac{\e}{1+\e^2
x^2}} .\] Using l'H\^opital's rule, we get $\lim_{x\r -\infty}
g(x,\e)=0$, $\lim_{\e \r 0} g(x,\e)=0$ for any $x<0$, and
\[ \lim_{x\r 0}g(x,\e )=\frac{\pi}{2u_c} \frac{-\displaystyle\int_R K(y)
\bigl[ \dfrac{\e}{1+\e^2 (x+y)^2}-\dfrac{\e}{1+\e^2 x^2} \bigr]
dy}{\dfrac{\e}{1+\e^2 x^2}\dfrac{\e}{1+\e^2 x^2} -{\rm tan}^{-1} (\e
x)\dfrac{2x \e^3}{(1+\e^2 x^2 )^2}} = \frac{\pi \e}{2u_c} \int_R
\dfrac{y^2 K(y)}{1+\e^2 y^2}dy.\] Since $\int_R y^2 K(y)dy <\infty$,
we see that for small enough $\e >0$, $g(x,\e )\leq 1$ for $x<0$,
therefore $s$ satisfies
\[ s+ss' \leq K*s ,\]
thus $s$ is a subsolution on $(-\infty,0)$.

On $(-\infty,0)$ define the sequence $\{ u_n \}_{n\geq 0}$ by
\begin{equation}
u_{n+1} +u_n u_{n+1}' =K*u_n \label{is}
\end{equation}
and $u_n (-\infty)=u_c$. This iteration has the same properties on
$(-\infty,0)$ and $(0,\infty)$, by odd extension and the evenness of
$K$. In the inductive step we show that the statement $\exists$
continuous $u_n$ such that $u_n
>0$, $u_n (-\infty)=u_c$, $u_n '\leq 0$, $u_n -u_{n-1}\leq 0$, $u_0
\geq u_n \geq s$ implies the statement $\exists$ continuous
$u_{n+1}$ such that $u_{n+1} >0$, $u_{n+1} (-\infty)=u_c$, $u_{n+1}
'\leq 0$, $u_{n+1} -u_n \leq 0$, $u_0 \geq u_{n+1} \geq s$.

The solution to the linear equation (\ref{is}) with the initial
condition $u_{n+1} (0)=u_{n+1}^0 >0$ and $x<0$ is given by the
formula
\begin{equation}
u_{n+1} (x)=e^{\int^0_x \frac{ds}{u_n (s)}} \bigl[ u_{n+1}^0 -
\int^0_x e^{-\int^0_s \frac{dr}{u_n (r)}} \frac{K*u_n (s)}{u_n
(s)}ds \bigr], \label{int}
\end{equation}
However, $u_{n+1}$ in (\ref{int}) is such that $u_{n+1}
(-\infty)=u_c$ only if we choose
\begin{equation} u_{n+1}^0
=\int^0_{-\infty} e^{-\int^0_s \frac{dr}{u_n (r)}} \frac{K*u_n
(s)}{u_n (s)}ds, \label{u0}
\end{equation}
which follows from l'H\^opital's rule:
\[ \lim_{x \r -\infty} \frac{u_{n+1}^0 -
\displaystyle\int^0_x e^{-\int^0_s \frac{dr}{u_n (r)}} \dfrac{K*u_n
(s)}{u_n (s)}ds}{e^{\int^0_x \frac{ds}{u_n (s)}}} =\lim_{x \r
-\infty} K*u_n (x)=u_c.\] Note that the integral in (\ref{u0}) is
convergent, since $u_n \leq u_c$ implies $e^{-\int^0_s \frac{dr}{u_n
(r)}} \leq e^{\frac{s}{u_c}}$. Let $v_{n+1}=u_{n+1}-u_n$, $v_n =u_n
-u_{n-1}$. From (\ref{is}) we get
\[ v_{n+1} +u_n u_{n+1}' -u_{n-1} u_n' =K*v_n .\]
After adding and subtracting $u_n u_n '$ on the left side we get
\begin{equation}
v_{n+1} +u_n v_{n+1}' +u_n ' v_n =K*v_n .\label{v}
\end{equation}
Since $v_n$ is odd, $K*v_n$ can be represented on $(-\infty,0)$ as
\begin{equation}
K*v_n (x)=\int_{-\infty}^{0} [K(x-y)-K(x+y)]v_n (y)dy . \label{r}
\end{equation}
Notice that $K(x-y)-K(x+y) \geq 0$ for $x\leq 0$ and $y\leq 0$
because $K$ is even and non-decreasing for $x<0$, thus $K*v_n \geq
0$. If $v_{n+1} >0$ somewhere, then since $v_{n+1} (-\infty)= 0$,
$\exists x_0$ such that $v_{n+1}(x_0 )>0$ and $v_{n+1}'(x_0 )>0$.
Since $v_n \leq 0$, $u_n '\leq 0$, the left side of (\ref{v}) is
positive, a contradiction. Now differentiate (\ref{is}):
\begin{equation}
u_{n+1}' +u_n 'u_{n+1}'+u_n u_{n+1}''=K'*u_n =-K(x)[u_n (0-) -u_n
(0+)]+K*u_n ' \leq 0 .\label{d}
\end{equation}
If $u_{n+1}'>0$ somewhere, then since $u_{n+1}\leq u_n$, $\exists
x_0$ such that $u_{n+1}' (x_0 )>0$ and $u_{n+1}'' (x_0 )>0$. The
left side of (\ref{d}) at $x_0$ is then positive since $u_n '(x_0
)>-1$ from (\ref{is}), a contradiction. Using $u_{n+1}'\leq 0$ we
get $u_{n+1} \geq u_{n+1} +u_n u_{n+1}'=K*u_n
>0$ for $x<0$.

Assuming $u_n \leq u_0$ we get
\[ u_{n+1} +u_0 u_{n+1}' \leq u_{n+1}+u_n u_{n+1}' =K*u_n \leq K*u_0
=u_1 +u_0 u_1 ', \] thus
\[ u_{n+1} -u_1 +u_0 (u_{n+1}'-u_1 ')\leq 0, \]
and since $u_{n+1}(-\infty)=u_0(-\infty)$, $u_{n+1}\leq u_0$ by a
similar argument as above. Assuming $u_n \geq s$, in the same way we
get
\[ u_{n+1} -u_1 +s (u_{n+1}' -u_1 ') \geq 0 \]
and $u_{n+1}\geq s$.

Thus $\{ u_n \}$ is a sequence of monotone functions such that
$u_{n+1} (x)\leq u_n (x)$ for all $n\geq 0$ and $x<0$. Let $u\equiv
\lim_{n\r \infty} u_n$. Since $u_0 \geq u_n \geq s$ for all $n\geq
0$, $u(-\infty )=u_c$. In (\ref{is}) add and subtract
$u_{n+1}u_{n+1}'$, multiply it by $\phi \in C^\infty$, with $\phi$
compactly supported in $(-\infty,0)$, and integrate over
$(-\infty,0)$:
\begin{equation}
\int_{-\infty}^0 \bigl[ -\frac{1}{2} u_{n+1}^2 \phi'+u_{n+1} \phi
-u_{n+1}' (u_{n+1}-u_n )\phi - (K*u_n )\phi \bigr]=0.\label{wl}
\end{equation}
From $u_0 \geq u_n \geq s$ and (\ref{is}), $|u_n '|$ is
uniformly bounded in $n$, thus we can use the Lebesgue Dominated
Convergence Theorem in (\ref{wl}) and pass to the limit as $n\r
\infty$ to get
\[ \int_{-\infty}^0 \bigl[ -\frac{1}{2} u^2 \phi' +(u-K*u)\phi
\bigr] =0.\] Since $\frac{1}{2} u^2 \in W^{1,\infty}(-\infty ,0)$,
$\frac{1}{2}u^2$ is differentiable a.e. and $(\frac{1}{2}
u^2)'=K*u-u$ a.e. on $(-\infty,0)$. $\frac{1}{2} u^2 \in
W^{1,\infty}(-\infty ,0)$ also implies that $\frac{1}{2}u^2$ and
therefore $u$ is Lipschitz continuous, thus $u\in C^1 (-\infty,0)$
and $u$ satifies (\ref{e}) on $(-\infty,0)$. Since $u$ is odd, it
also satisfies (\ref{e}) on $R \setminus 0$ and (\ref{wf}).
$U'(x)<0$ for $x\neq 0$ is shown in a similar way as the argument
below (\ref{d}).

To find a condition for which $u$ is discontinuous, we argue by
contradiction. Assume that $u$ is continuous at $0$. Note that $u'$
is integrable on $R$, since for $x\neq 0$ we have
$u'=\frac{K*u}{u}-1\geq -1$. Integrating (\ref{e}) over
$(-\infty,0)$ and using Fubini's Theorem and then the LDC Theorem we
get
\[ -\frac{1}{2}{u_c}^2 = \int_{-\infty}^0 uu' =
\int_{-\infty}^0 (K*u-u)  =\int_{-\infty}^0 \int_R K(y)\int_0^1
yu'(x+yt)dtdydx \] \[= \int_R yK(y) \int_0^1 \lim_{m\r
-\infty}[u(yt)-u(-m+yt)]dtdy = \int_R yK(y) \int_0^1 u(yt)dtdy .\]
The right side can be estimated to yield
\[ \frac{1}{2} {u_c}^2 \leq u_c \int_R |y|K(y)dy .\]
Thus if ${u_c}^2 >2 \int_R |y|K(y)dy$ we obtain a contradiction.
This gives the criterion for a solution to be discontinuous as $u_-
-u_+ >4 \int_R |y|K(y)dy$, since $u_c =\frac{1}{2} (u_- -u_+ )$. For
$K(y)=\frac{k}{2}e^{-k|y|}$, we get $u_- -u_+ > \frac{4}{k}$. \ep

\bibliographystyle{unsrt}

\end{document}